\documentclass[twocolumn,aps,showpacs]{revtex4}
\usepackage{graphicx}
\usepackage{dcolumn}
\begin{document}

\title{Voltage Control of Exchange Coupling in Phosphorus Doped Silicon}
\author{C.J. Wellard$^a$, L.C.L. Hollenberg$^a$, L.M. Kettle$^{b}$ and  H.-S. Goan$^{c}$}
\affiliation{Centre for Quantum Computer Technology,\\
$^a$ School of Physics, University of
Melbourne, Victoria 3010, AUSTRALIA. \\
$^b$ University of Queensland, QLD 4072, AUSTRALIA \\
$^c$ University of New South Wales, Sydney NSW 2052, AUSTRALIA.
}

\date{\today}

\begin{abstract}
Motivated by applications to quantum computer architectures we study the change in the exchange interaction between neighbouring phosphorus donor electrons in silicon due to the application of voltage biases to surface control electrodes. These voltage biases create electro-static fields within the crystal substrate, perturbing the states of the donor electrons and thus altering the strength of the exchange interaction between them. We find that control gates of this kind can be used to either enhance, or reduce the strength of the interaction, by an amount that depends both on the magnitude and orientation of the donor separation.

\end{abstract}

\pacs{ 03.67.Lx, 71.55.Cn, 85.30.De}

\maketitle

\section{introduction}
\label{section:intro}
Phosphorus donors in silicon have been the subject of increased research interest in recent years due to their status as either nuclear \cite{Kane98}, or electron \cite{Vrijen00} -spin qubits in various proposals for a scalable quantum computer. In both the cases the exchange energy between neighbouring donor electrons is of fundamental importance as the mediator of the the qubit coupling. Although proposals exist for implementing quantum information processing with this exchange interaction fixed \cite{Benjamin}, it is none the less desirable to have some control over this parameter, particularly for the readout process. Recent studies have shown that the magnitude of the exchange interaction for donor electrons in silicon is strongly dependent on not only the magnitude of the donor separation, but also on the relative orientation of the donors within the silicon crystal lattice \cite{Koiller02,Koiller02A,Wellard03}.
\par
\begin{figure}[h]
\includegraphics[angle=0, width=0.5\textwidth]{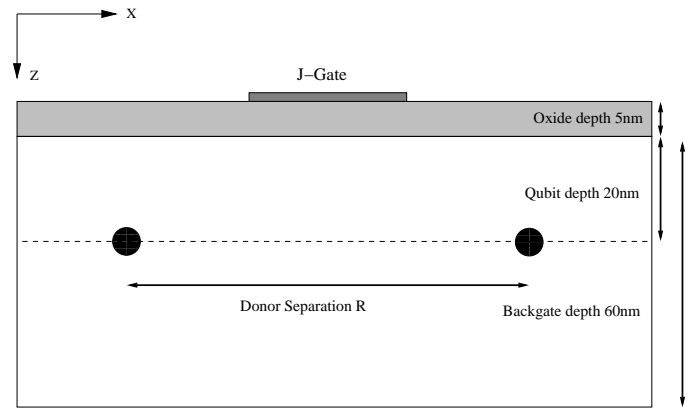}
\caption{\label{fig:kanepic}The Kane architecture based on buried
phosphorus dopants in a silicon substrate}
\end{figure}

 In a previous article \cite{Wellard03} the effect of J-gate biases on the exchange coupling was computed for donor separations along the [100] axis only. Here we extend the calculations to include separations in other orientations relative to the host silicon lattice, in particular separations along the [110] and [111] crystallographic axes. A two-qubit Kane device, Fig.\ref{fig:kanepic}, consists of two phosphorus donors at substitutional sites in a silicon substrate at a depth $20$nm  below a $50$nm layer of silicon oxide. On top of this oxide, and between the donors is a metallic J-gate electrode, which, for the purposes of these calculations is assumed to be $7$nm  wide, and infinitely long. A grounded back plane lies at a depth of $600 $\AA below the silicon-oxide interface. The distance $R$, between the donors is varied as is the orientations relative to the silicon crystallographic axis. The electro-static potential created inside the device by the application of a voltage bias to the J-gate electrode, while the remaining electrodes are held at ground,  is calculated by means of a commercial package that solves the Poisson equation for such semiconductor systems \cite{TCAD}, details of these calculations for Kane type devices can be found in ref \cite{Pakes02}.

\section{Donor electron wave-functions}
\label{section:wavefuns}
In the absence of a perturbing potential the electron wave-function for a phosphorus donor centred at a position ${\bf R}/2$, in a silicon substrate can be expressed in the Kohn-Luttinger \cite{Kohn55,Kohn57} form:
\begin{equation}
\psi({\bf r-R}/2) = \sum_{\mu} F^{1,0,0}_\mu({\bf r-R}/2) {\rm e}^{\bf k_\mu
.(r-R) } u_\mu ({\bf r}).
 \label{equation:donor_wf}
\end{equation}
The functions $u_\mu ({\bf r})$ are the periodic part of the Bloch functions for the pure silicon crystal, with wave-vectors ${\bf k}_\mu$ located at each of the degenerate conduction band minima. The envelope functions ,
\begin{equation}
F^{1,0,0}_{\pm z}({\bf r}) = \varphi_{1,0,0}(x,y,\gamma z),
\label{equation:envelope}
\end{equation}
are correctly normalised non-isotropic hydrogen like wave-functions, with effective Bohr radius $a_\perp$. The above example is for functions localised around the conduction band minima  ${\bf k}_{\pm z} = 2 \pi/a (0,0,\pm 0.85)$, with $a=5.43$\AA \ the lattice spacing for silicon. The parameter $\gamma = a_\perp/a_\parallel$, where the values $a_\perp = 25.09 {\rm \AA},a_\parallel = 14.43 {\rm \AA}$, are transverse and longitudinal effective Bohr radii respectively, and are determined variationally \cite{Kohn55,Koiller02}. The subscripts refer to the electronic, orbital and magnetic $(n,l,m)$ quantum numbers respectively.
\par To control the strength of the exchange coupling it is proposed that a voltage bias be applied to a surface ``J-gate'' electrode, see Fig.\ref{fig:kanepic}. This voltage bias produces an electric potential within the substrate, perturbing the donor electron wave-functions, and thus altering the exchange coupling. To model this process we have calculated the electro-static potential produced inside the device by the application of a J-gate potential. The potential matrix is then calculated in the basis of generalised Kohn-Luttinger states given in Eq.\ref{equation:donor_wf}, using hydrogenic envelope functions up to and including $n=7$, a basis of 140 states in total. The Hamiltonian is then diagonalised in this basis to find the ground-state donor electron wave-function:
\begin{eqnarray}
\psi({\bf r-R}/2;V) &=& \sum_{n,l,m} c_{n,l,m}(V) \\
& \times & \sum_{\mu} F_\mu^{n,l,m}({\bf r-R}/2) {\rm e}^{\bf k_\mu
.(r-R)/2 } u_\mu ({\bf r}).\nonumber
\label{equation:donor_wf(v)}
\end{eqnarray}
\par In the Heitler-London approximation the two donor system is treated as a correctly symmetrised product of single donor wave-functions. The fermion anti-symmetry of the overall wave-function ensures that the spin singlet (s), and triplet (t) states are respectively represented as even and odd superpositions of single electron wave-functions. This approximation is valid when the donor separation is large compared to the effective Bohr radii of the single electron wave-functions. For donors centred at positions $\pm {\bf R}$, the spatial part of the triplet/singlet wave-functions are thus:
\begin{eqnarray}
\Psi_{t/s}({\bf r}_1,{\bf r}_2;V) &=& \frac{1}{\sqrt 2} \{ \psi({\bf r}_1 -{\bf R}/2;V) \psi({\bf r}_2 +{\bf R}/2;V) \nonumber\\ 
& \pm & \psi({\bf r}_1 +{\bf R}/2;V) \psi({\bf r}_2 -{\bf R}/2;V) \}.\nonumber\\
\label{equation:psipm}
\end{eqnarray}
\section{Exchange Coupling}
\label{section:exchange}
In the Kane proposal for a phosphorus nuclear spin quantum computer in silicon \cite{Kane98}, the effective Hamiltonian between neighbouring donors is given by
\begin{eqnarray}
H_{eff} &=& g_n \mu_n B (\sigma_{n1}^z + \sigma_{n2}^z) +  g \mu_B B (\sigma_{e1}^z + \sigma_{e2}^z) \\
&+& A_1(V) {\vec \sigma}_{n1} \cdot {\vec \sigma}_{e1} + A_2(V) {\vec \sigma}_{n2} \cdot {\vec \sigma}_{e2} + J(V) {\vec \sigma}_{e1}\cdot{\vec \sigma}_{e2}. \nonumber
\label{equation:Heff}
\end{eqnarray}
Here the $\sigma$ represent the usual Pauli operators, and the subscripts $n,e$ represent operation on nuclei and electrons respectively. The controllable parameters, through which quantum information processing is implemented, are the contact hyperfine coupling $A$, between the donor nucleus and its associated electron, and the exchange coupling $J$ between neighbouring donor electrons. Both these quantities are to be controlled through manipulation of the electron wavefunction via the application of bias voltages to control gates. The contact hyperfine coupling is dependent on $|\psi(0)|^2$, that is, the probability density of the electron wave-function located at the position of the nucleus. This can be changed by drawing the electron wave-function away from the nucleus via the application of a positive bias to a A-gate \cite{Wellard02,Kettle03}. 
The exchange coupling, as can be seen from Eq.\ref{equation:Heff}, is equal to a quarter of the energy difference between the two electron spin single, and spin triplet states. In the Heitler-London approach, this is evaluated by calculating the energy difference between the even and odd superpositions Eq.\ref{equation:psipm}. The origin of this difference in energy between the spin singlet and triplet sates is due to the fact that the even superposition state has a higher probability density in the region located between the two donors, a region that has a relatively low potential. The odd superposition state has a lower probability density in this region which is compensated by a higher probability density in the higher potential regions located outside the two donors. This energy difference can be manipulated by application of a voltage bias to a J-gate located above and between the two donors. Application of a positive bias will decrease the potential in the region between the two donors, thus increasing the exchange coupling while a negative bias will have the opposite effect. 
\section{Results}
\label{section:results}
It has been well documented that in the absence of a bias potential, the exchange coupling between neighbouring phosphorus donor electrons in a silicon substrate is strongly dependent on both the magnitude and the orientation of the donor separation with respect to the host silicon lattice \cite{Koiller02,Wellard03}. 
\par
\begin{figure}
\rotatebox{-90}{\resizebox{5cm}{!}{\includegraphics{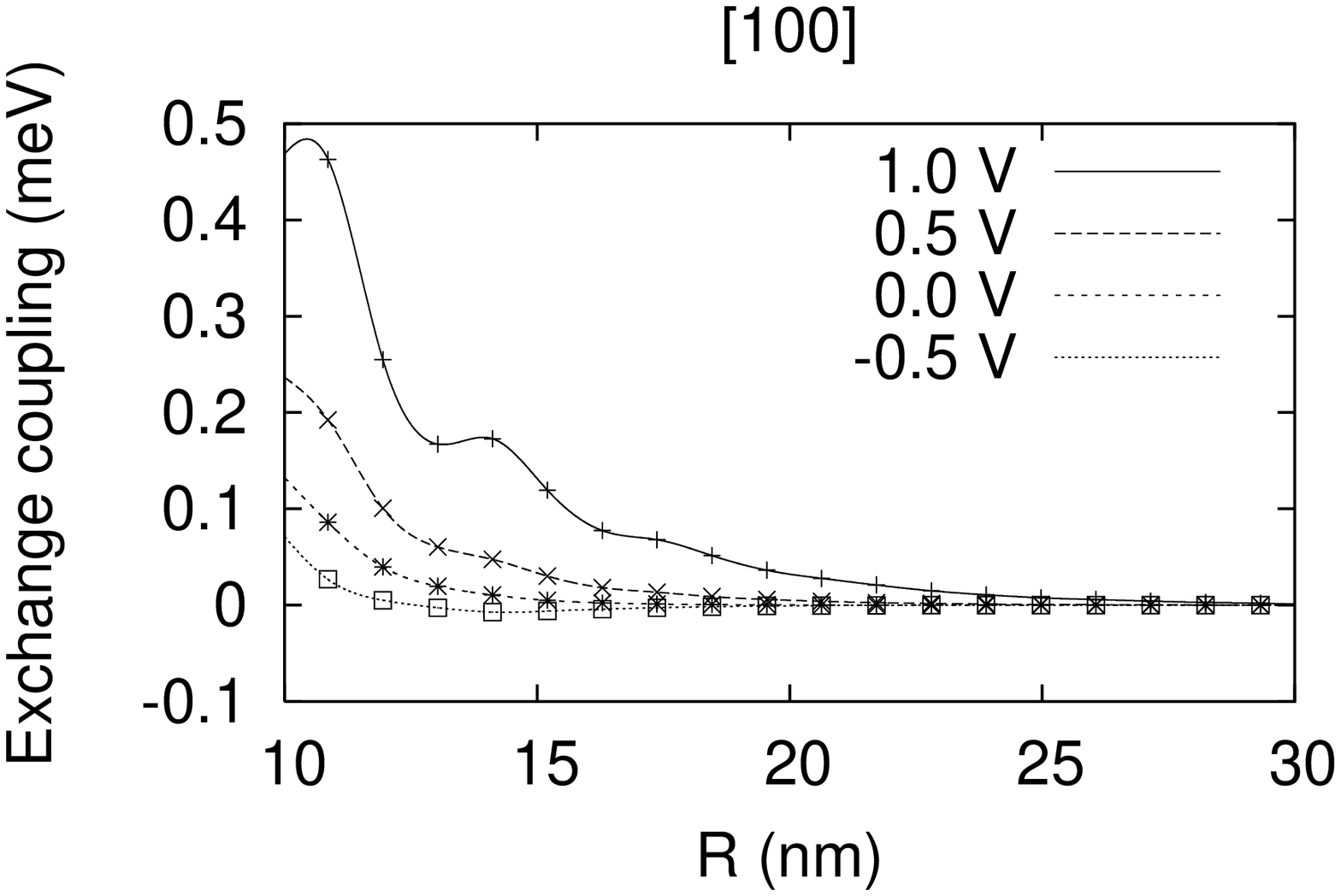}}}
\rotatebox{-90}{\resizebox{5cm}{!}{\includegraphics{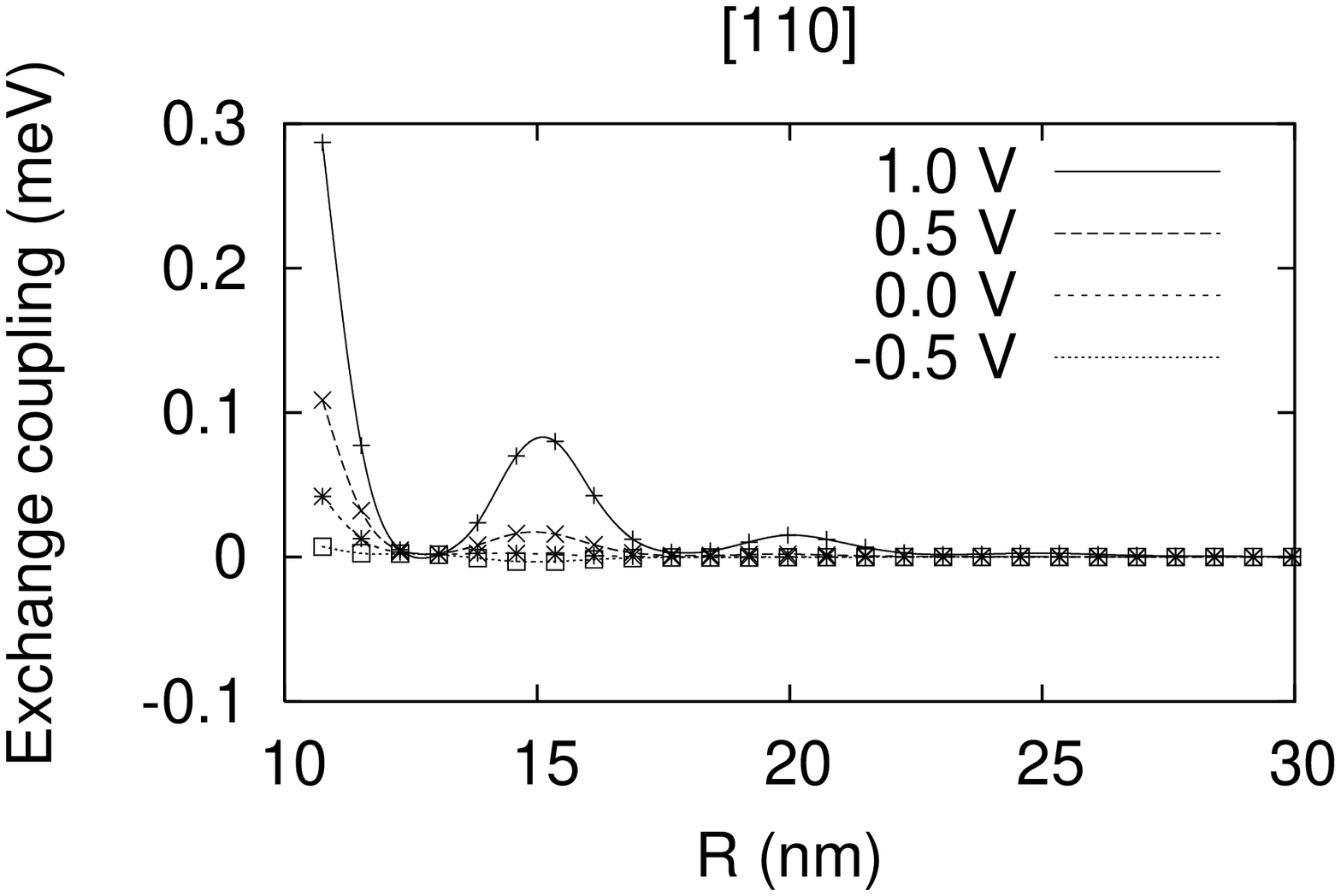}}}
\rotatebox{-90}{\resizebox{5cm}{!}{\includegraphics{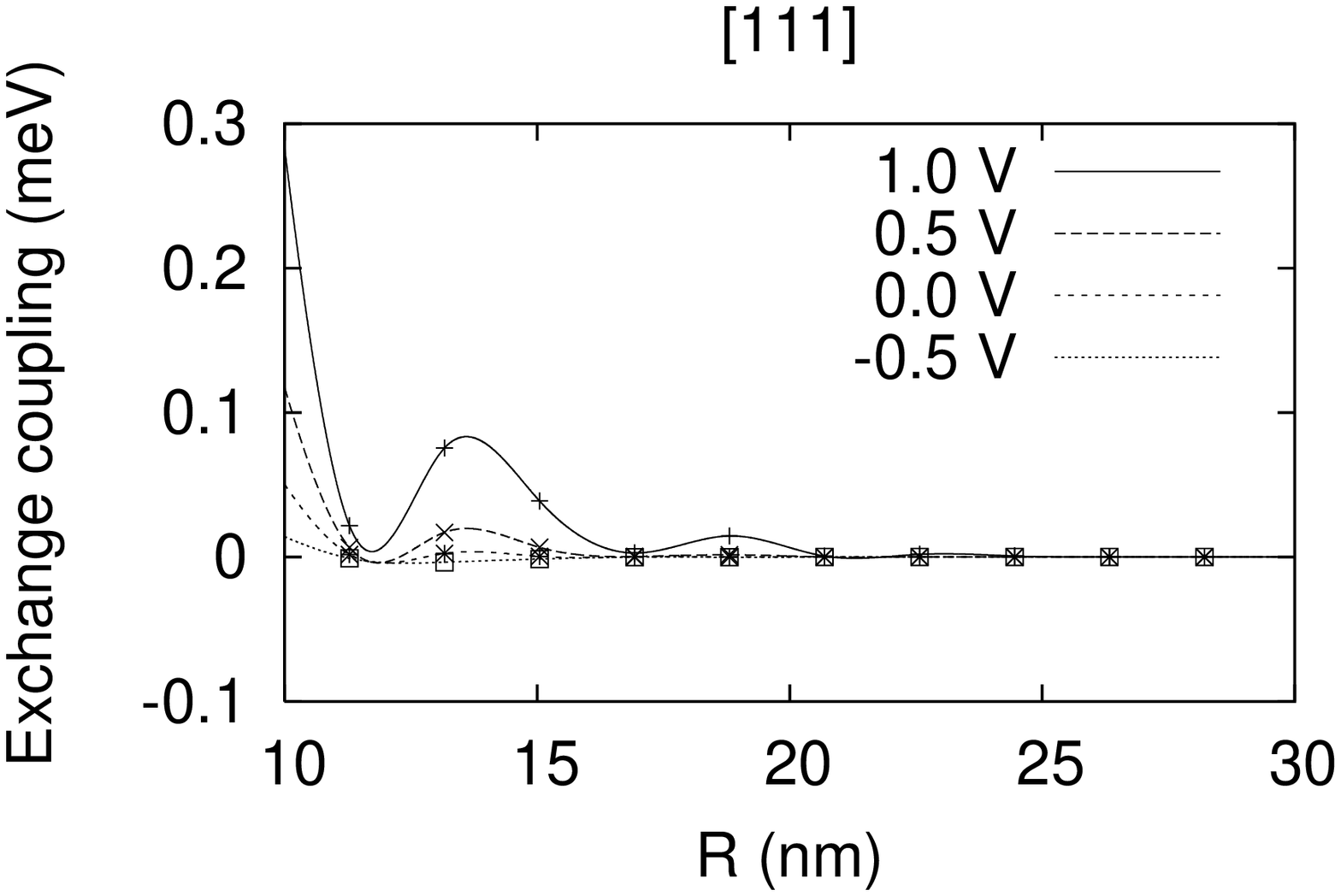}}}
\caption{Exchange coupling as a function of donor separation along high symmetry crystal directions for different J-gate biases. The points represent the position of fcc substitutional sites.}
\label{fig:vdep}
\end{figure}

 With this in mind it is not unexpected that this orientation dependence remain in the presence of a bias potential, Indeed we find that the presence of this potential enhances the oscillations observed in the zero bias case. This is illustrated in Fig.\ref{fig:vdep}, for separations of increasing magnitude along three high symmetry crystallographic axes.

\begin{figure}
\rotatebox{-90}{\resizebox{5cm}{!}{\includegraphics{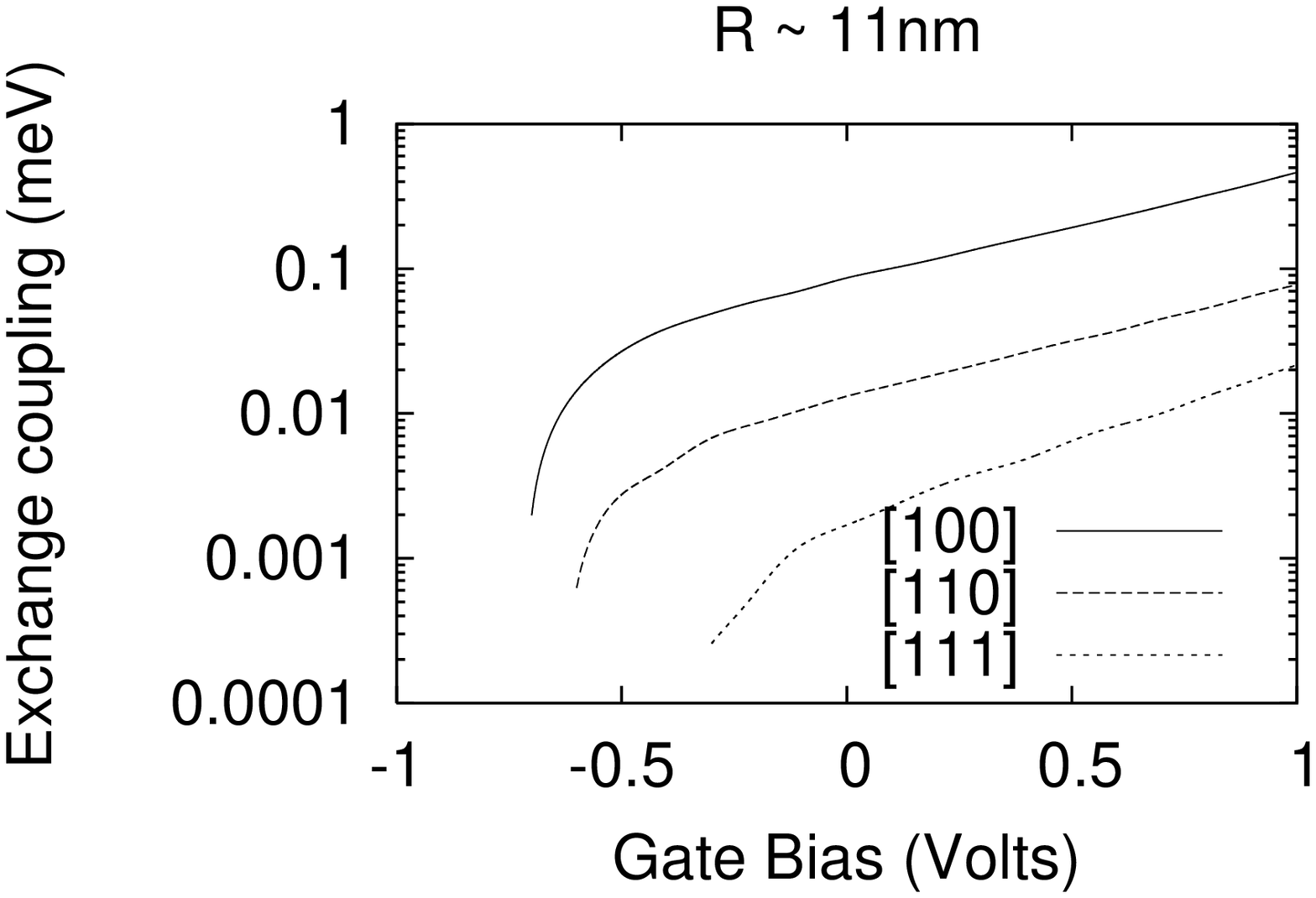}}}
\rotatebox{-90}{\resizebox{5cm}{!}{\includegraphics{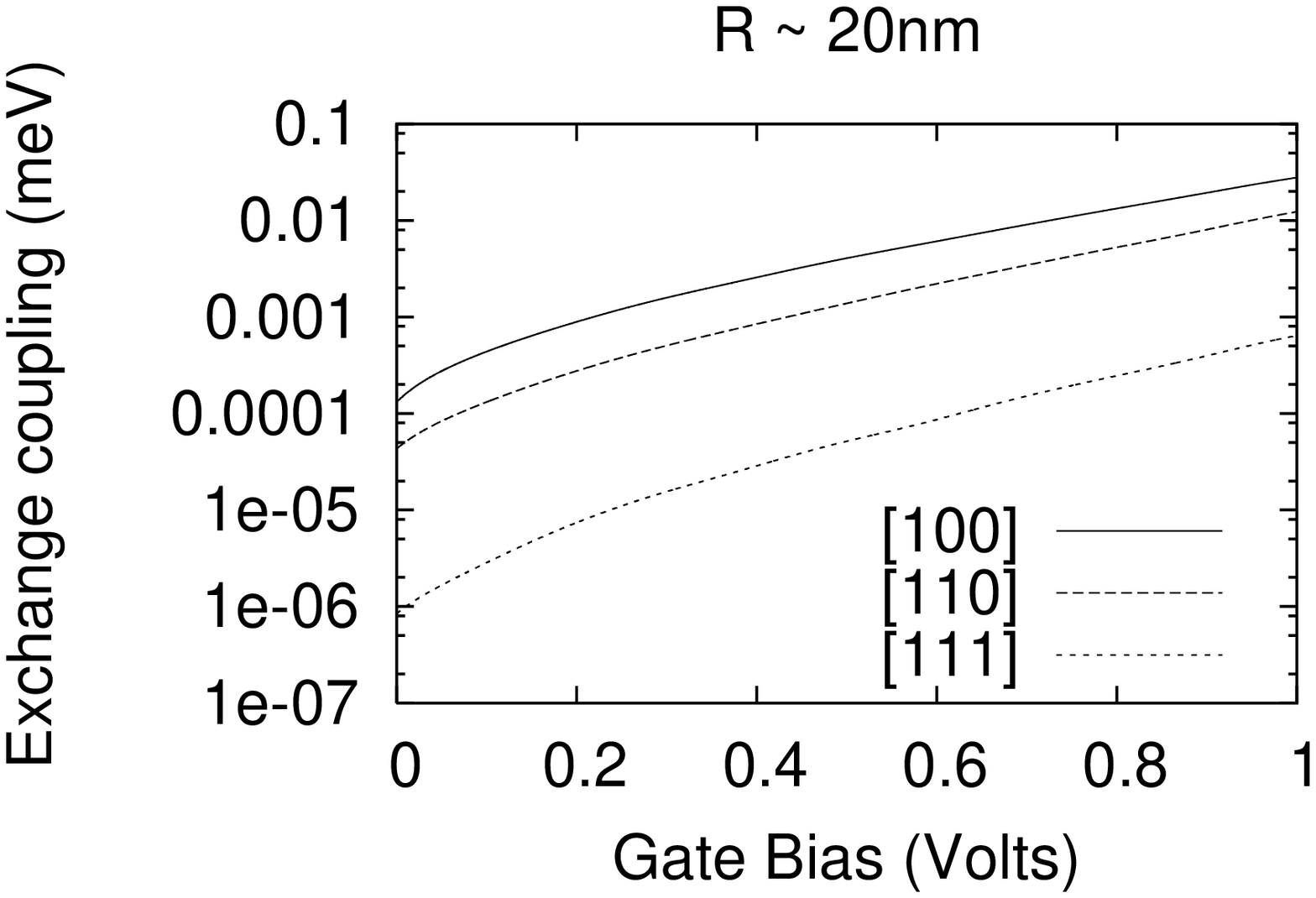}}}
\caption{Semi-Log plot of the exchange coupling as a function of applied gate bias along high symmetry crystal axes for different separation magnitudes. Because of the discrete positioning of possible substitutional sites, the exact magnitude of the separation is different for each direction with $R=10.86 nm$ in the [100] direction, $R=11.51nm$ for the [110] separation and $R=11.28$ for the donor separations along the [111] axis.}
\label{fig:rdep}
\end{figure}

It is evident from these plots, that application of a voltage bias to a J-gate can exert significant control over the strength of the exchange coupling, both increasing and decreasing the strength depending on the polarity of the bias. The range over which the exchange coupling can be changed is however extremely sensitive to the donor separation. This is well illustrated in Fig.\ref{fig:rdep} where we have plotted the value of the exchange coupling as a function of applied voltage bias for donor separations of different magnitudes along the same crystallographic axes. 

For all separations studied it was found to be possible, via the application of a suitable negative voltage bias, to reduce the  exchange coupling to zero, and thus to turn the inter-qubit interaction off. Indeed calculations at large negative bias, particularly for large donor separations, give negative values of the exchange coupling, a result that is unphysical for two electron systems. We attribute this result to a breakdown of the Heitler-London approximation in this regime. 

The speed of two-qubit gates, for both electron and nuclear spin architectures increases with the strength of the exchange coupling and so it would be desirable to achieve the largest exchange coupling possible. To this end it is clear that a donor separation along the [100] axis would be preferable, with the magnitude of the separation being as small as possible. Of course there are limits on the precision to which these donors can be placed, and on the separations that can be achieved. Technical issues will probably set a lower bound on reliable donor separation of approximately $20$nm. Due to the process by which the silicon is doped in the so called ``Bottom up'' approach \cite{Schofield03}, it is possible to place donors with atomic precision in the same [001] plane (that is a plane perpendicular to a crystallographic [001] axis) reliably using atomic force microscopy techniques and thermal incorporation. During the process of silicon over growth however the phosphorus atoms are likely to diffuse within this plane by up to several nanometers, diffusion out of the plane is expected to be significantly less. 

With this in mind we have calculated the expected exchange coupling for donors located at various substitutional sites within the [001] plane, with a separation magnitude of approximately $20$nm for different donor orientations, the results are presented in Fig.\ref{fig:theta}. The first plot shows the exchange coupling for donors separated by a vector ${\vec R}_\theta = R (\cos \theta,\sin \theta , 0)$, that is the donors are in the same [001] plane, and when $\theta =0 $ are separated by vector in the [100] direction. The second and third plot shows the strength of the exchange coupling when the two donors are not in the same [001] plane. In the second plot one of the donors has been displaced to its nearest neighbour substitutional site and the donor separation is now ${\vec R} = {\vec R}_\theta + a/4(1,1,1)$. In the third plot the donor is displaced by one lattice constant perpendicular to the [001] plane, ${\vec R} = {\vec R}_\theta + a(0,0,1)$. The plots show that even a small displacements from the ideal [100] separation significantly decrease the strength of the exchange coupling, indeed a displacement of just one lattice site leads to a reduction of approximately one half. 

\begin{figure}
\rotatebox{0}{\resizebox{6cm}{!}{\includegraphics{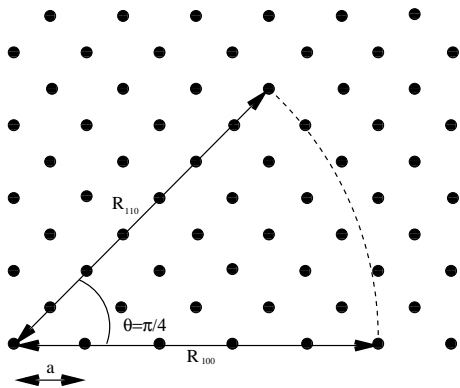}}}
\caption{\label{fig:plane} Schematic of the crystallographic structure of a [001] plane in silicon, showing donors separated by vectors along a [100] axis (${\bf R}_{100}$), and a [110] axis (${\bf R}_{110}$). For two donors in the same [001] plane, we can define the separation $R_\theta$ relative to the [100] axis.}
\end{figure}

\begin{figure}
\rotatebox{-90}{\resizebox{5cm}{!}{\includegraphics{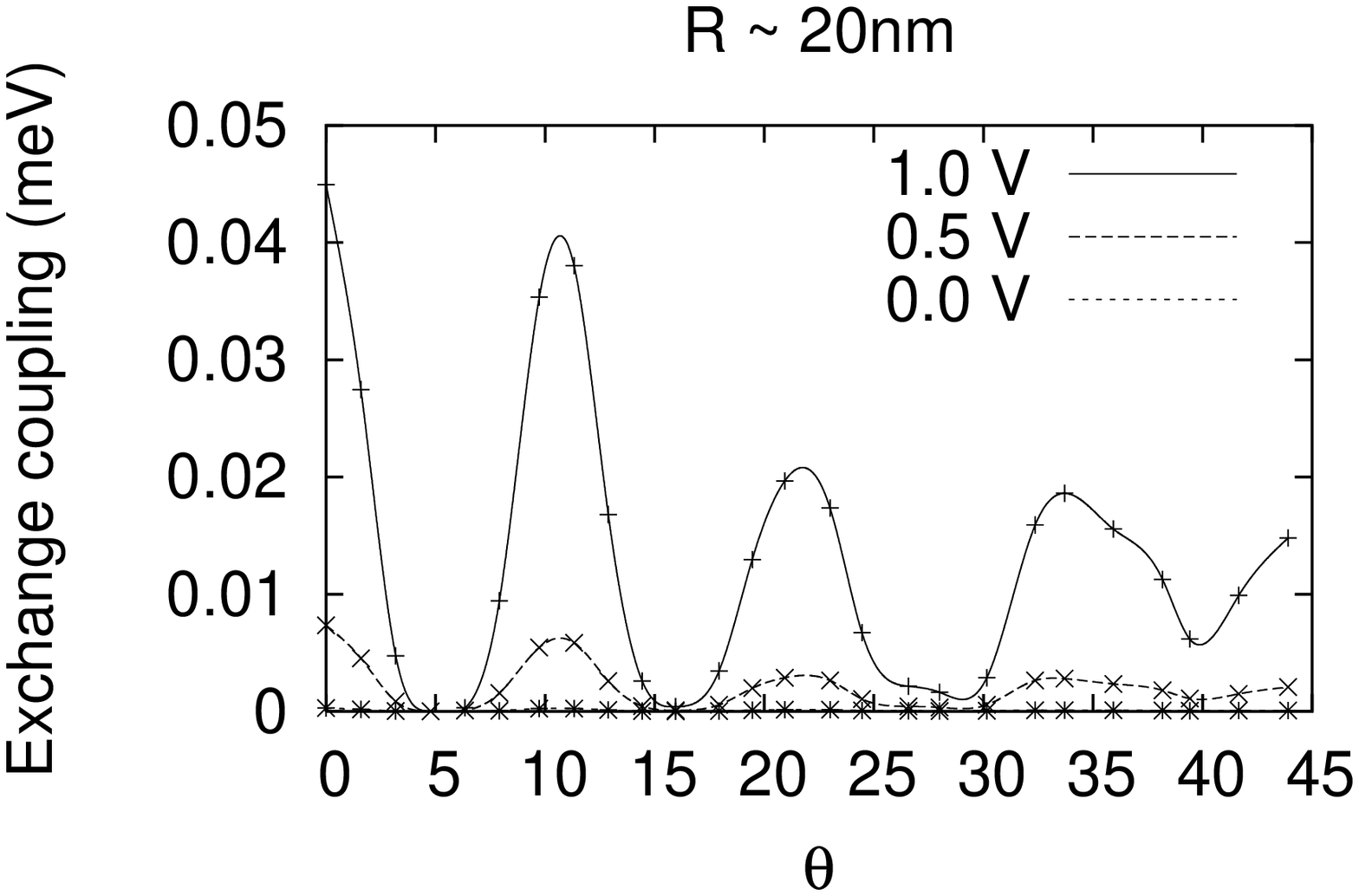}}}
\rotatebox{-90}{\resizebox{5cm}{!}{\includegraphics{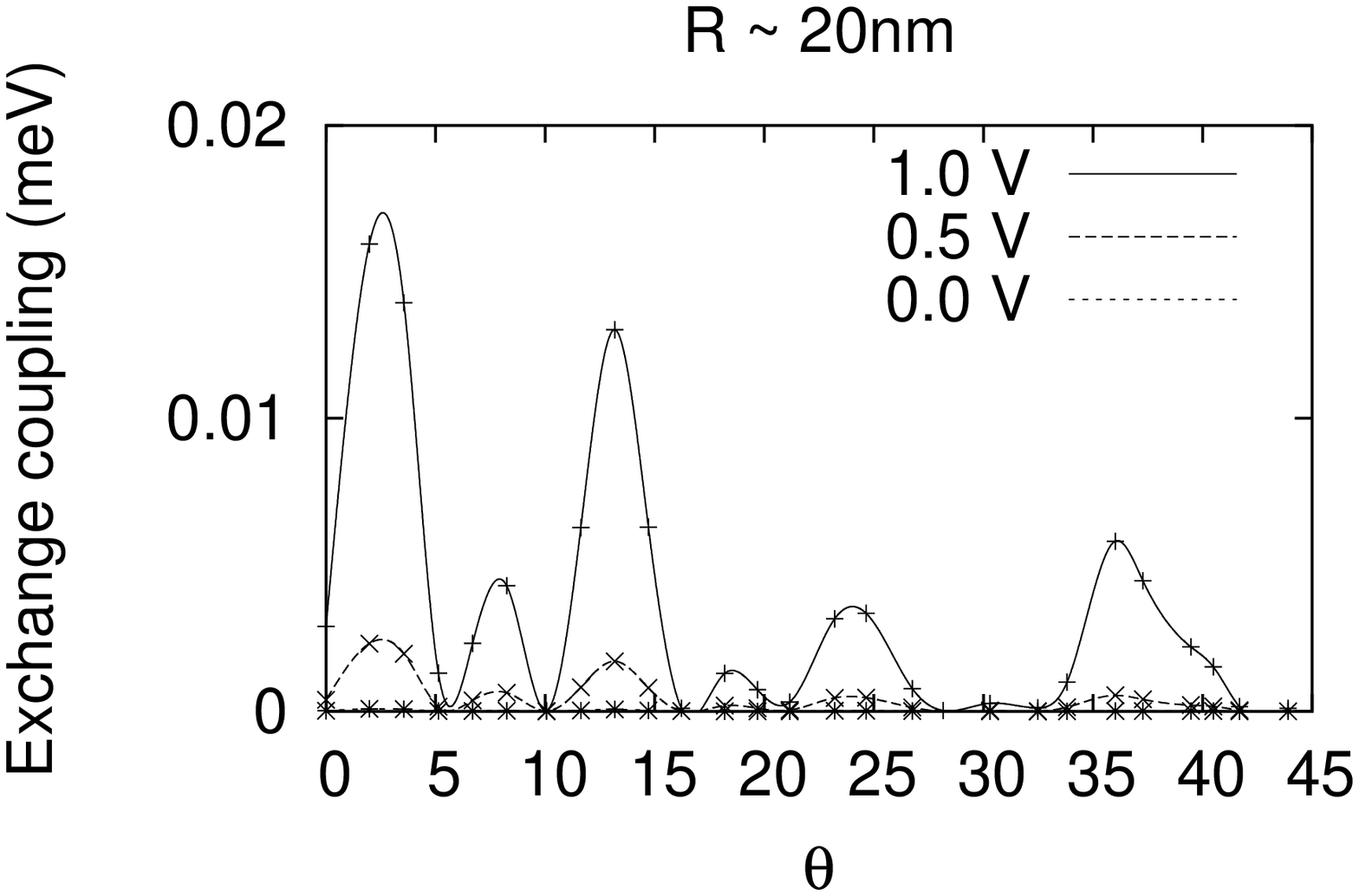}}}
\rotatebox{-90}{\resizebox{5cm}{!}{\includegraphics{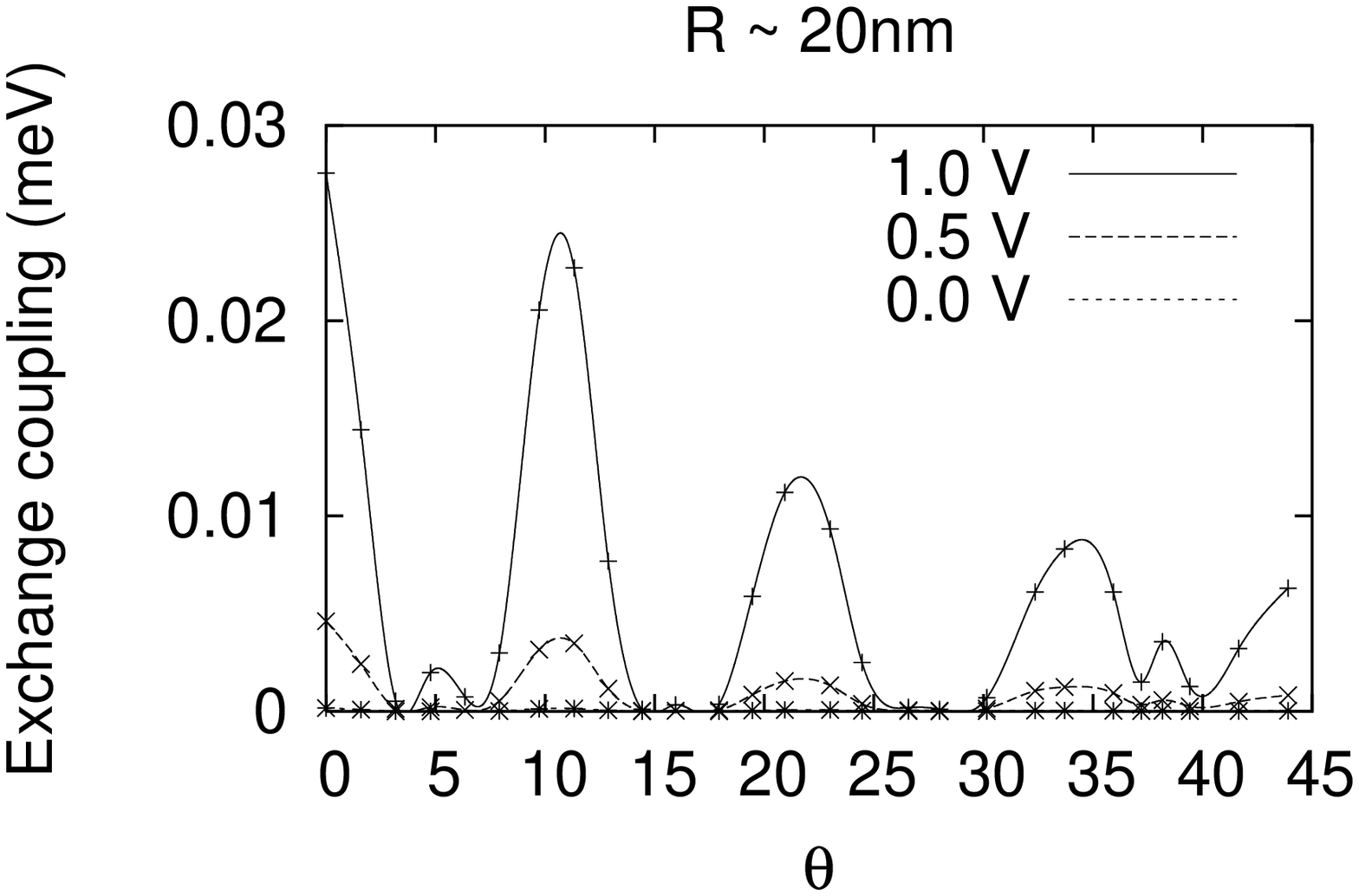}}}
\caption{Exchange coupling as a function of donor separation angle from the [100] direction for donors separated by approximately $200$ \AA \ in various planes. In the first figure both donors are in the same [100] plane, and the angle $\theta$ refers to the donor orientation with respect to separation along the [100] axis,  as shown in Fig.\ref{fig:plane}. In the second figure one donor has been moved to a nearest-neighbour site, a displacement of $\delta = a/4(1,1,1)$ relative to the positions of the first plot. The final plot contains data for which the second donor has been displaced by one lattice constant in a direction out of the plane, $\delta = a(0,0,1)$. Here the points represent the position of substitutional fcc sites. The exact magnitude of the donor separation at these points varies slightly due to the discrete distribution of such sites.}
\label{fig:theta}
\end{figure}

\section{Conclusions}
We find that strength of the exchange coupling between neighbouring phosphorus donor electrons in a silicon substrate can be significantly increased or decreased by the application of a bias voltage to a surface ``J-gate'' electrode placed above and between the donor qubits. The bare (zero-bias) coupling is strongly dependent on the magnitude and orientation of the donor separation, and this dependence is amplified by the application of the voltage bias which has the affect of altering the coupling strength by a voltage dependent factor. Regardless of the donor orientation the exchange coupling strength can be reduced to zero, via the application of a negative J-gate bias, and so ideal donor placement is determined by the maximum value to which the coupling can be increased. To maximise the achievable coupling strength the donors would ideally be separated along a crystallographic [100] axis, however even small deviations from this ideal placement, of the order of one lattice spacing, can decrease the coupling strength by more than a factor of two.

\section{Acknowledgements} 
This work was supported by the
Australian Research Council, computational support was provided by the Australian Partnership for
Advanced Computing, as well as the Victorian Partnership for Advanced Computing. HSG would like to
acknowledge the support of a Hewlett-Packard fellowship.

\end{document}